\documentclass{ieeetj}
\usepackage{cite}
\usepackage{amsmath,amssymb,amsfonts}
\usepackage{algorithmic}
\usepackage{graphicx,color}
\usepackage{textcomp}
\usepackage{xcolor}
\usepackage{hyperref}
\usepackage{algorithm,algorithmic}
\usepackage[most]{tcolorbox}
\usepackage{tikz}
\usepackage{etoolbox}
\usepackage{stfloats} 
\usepackage{graphicx}
\usepackage{subcaption} 
\usepackage{booktabs} 
\usepackage{array}    
\def\BibTeX{{\rm B\kern-.05em{\sc i\kern-.025em b}\kern-.08em
    T\kern-.1667em\lower.7ex\hbox{E}\kern-.125emX}}
\AtBeginDocument{\definecolor{tmlcncolor}{cmyk}{0.93,0.59,0.15,0.02}\definecolor{NavyBlue}{RGB}{0,86,125}}

\def\OJlogo{\vspace{-4pt}}
\def\seclogo{\vspace{10pt}}

\def\authorrefmark#1{\ensuremath{^{\textbf{#1}}}}

\begin{document}
\markboth{}{PATWARI {ET AL.} Discovering Optimal Robust Minimum Redundancy Arrays (RMRAs) through Exhaustive Search}

\title{Discovering Optimal Robust Minimum Redundancy Arrays (RMRAs) through Exhaustive Search and Algebraic Formulation of a New Sub-Optimal RMRA}

\author{Ashish Patwari\authorrefmark{1}, Member, IEEE, Sanjeeva Reddy S\authorrefmark{1}, Student Member, IEEE \\ and G Ramachandra Reddy\authorrefmark{1}\, Senior Member, IEEE}
\affil{School of Electronics Engineering, Vellore Institute of Technology, Vellore, Tamil Nadu, India.}
\corresp{Corresponding author: Ashish Patwari (Email: ashish.p@vit.ac.in).}

\begin{abstract}
Modern sparse arrays are maximally economic in that they retain just as many sensors required to provide a specific aperture while maintaining a hole-free difference coarray. As a result, these are susceptible to the failure of even a single sensor. Contrarily, two-fold redundant sparse arrays (TFRSAs) and robust minimum redundancy arrays (RMRAs) ensure robustness against single-sensor failures due to their inherent redundancy in their coarrays. At present, optimal RMRA configurations are known only for arrays with sensor counts $N=6$ to $N=10$. To this end, this paper proposes two objectives: (i) developing a systematic algorithm to discover optimal RMRAs for $N>10$, and (ii) obtaining a new family of near-/sub-optimal RMRA that can be completely specified using closed-form expressions (CFEs). We solve the combinatorial optimization problem of finding RMRAs using an exhaustive search technique implemented in MATLAB.  Optimal RMRAs for \(N = 11\) to \(14\) were successfully found and near/sub-optimal arrays for \(N = 15\) to \(20\) were determined using the proposed technique. As a byproduct of the exhaustive search, a large catalogue of valid near- and sub-optimal RMRAs was also obtained. In the second stage, CFEs for a new TFRSA were obtained by applying pattern mining and algebraic generalizations to the arrays obtained through exhaustive search. The proposed family enjoys CFEs for sensor positions, available aperture, and achievable degrees of freedom (DOFs). The CFEs have been thoroughly validated using MATLAB and are found to be valid for $N\geq8$. Hence, it can be concluded that the novelty of this work is two-fold: extending the catalogue of known optimal RMRAs and formulating a sub-optimal RMRA that abides by CFEs.
\end{abstract}


\maketitle
\section{INTRODUCTION}
Sensor arrays are utilized in various fields, including radar, sonar, radio astronomy, medical imaging, wireless communications, and acoustic source localization. They consist of sensors—antennas, microphones, or hydrophones—in geometric layouts such as linear, planar, circular, or spherical. Their goal is to estimate parameters like the direction of arrival (DOA) of signals from multiple sources. Uniform Linear Arrays (ULAs), characterized by sensors evenly spaced along a line, represent the simplest and most prevalent design configuration. However, achieving large apertures and high resolution requires many sensors, increasing complexity and cost. 

Sparse linear arrays (SLAs) use fewer sensors than ULAs to achieve comparable aperture and angular resolution. Notably, SLAs can resolve more source angles than the number of physical sensors by exploiting second-order difference coarrays, which effectively increase the array’s degrees of freedom. The difference coarray (DCA) is the set of all pairwise differences between sensor positions and directly determines the spatial lags available for DOA estimation. Traditional ULAs can distinguish up to $N-1$ sources using $N$ elements \cite{re1}. In contrast, sparse arrays, including Minimum Redundancy Arrays (MRAs), nested arrays, and co-prime arrays, use second-order DCA processing, and can resolve up to \(O(N^2)\) sources. Co-prime arrays and two-level nested arrays were among the first sparse arrays with closed-form expressions (CFEs) for sensor placement, and obviated the need for exhaustive search algorithms. Since then, research into sparse array design has expanded rapidly. 

MRAs aim to achieve the largest aperture and a hole-free difference co-array, for a given number of sensors \cite{re2}. The nested array accomplishes a contiguous difference coarray with increased degrees of freedom (DOFs) through the combination of a dense subarray and a sparse subarray \cite{re3}. Conversely, the coprime array utilizes number-theoretic properties of coprime integers to establish two interleaved subarrays with sparse placement, resulting in a large aperture and a scalable design \cite{re4}. Collectively, these classical configurations offer substantial trade-offs among aperture, redundancy, and practical feasibility, serving as a foundation for ongoing investigations into resilient sparse array architectures. However, most sparse arrays are devised to maximally economic in terms of coarray/spatial redundancy, making them susceptible to the failure of even a single sensor. Moreover, MRAs encounter limited scalability due to combinatorial construction, nested arrays are susceptible to potential mutual coupling among closely spaced elements, and coprime arrays generate non-contiguous coarray structures that may necessitate coarray interpolation, a computationally complex task.

Enhancing coarray redundancy through the utilization of two-fold redundant sparse arrays (TFRSAs) can augment the robustness of the array against such failures. Nevertheless, the identification of novel TFRSAs remains a substantial challenge within the sparse array literature, principally due to the NP-hard nature of the problem, which entails determining optimal sensor placements that fulfill various constraints. 

Numerous TFRSA formulations have been proposed in existing scholarly works \cite{re5}, \cite{re6}, \cite{re7}, \cite{re8}, \cite{re9}. Liu and Vaidyanathan proposed RMRAs with the maximum aperture for specified sensors \((N)\) that satisfy two-fold redundancy \cite{re5}. However, they lacked CFEs and required exhaustive search methods, rendering larger arrays \((N>10)\) impractical. The configurations of optimal RMRAs for \(N>10\) remain unknown. SymNA is equipped with CFEs but only for \(N\geq16\) that are even \cite{re5}. The complex design of the composite Singer array limits its applicability to certain $N$ \cite{re8}, as does the fractal sparse array \cite{re9}. The array proposed by Zhu et al. has CFEs for all \(N \geq6\) \cite{re10}, utilising double difference bases (DDBs) derived from number theory as TFRSAs \cite{re6}. Nonetheless, not all DDBs are TFRSAs \cite{re11}; some possess hidden dependencies that may compromise difference coarray (DCA) continuity, thereby conflicting with the requirement of two-fold redundancy. Overall, it can be concluded that there are currently no TFRSA designs that (i) have CFEs, (ii) are defined for all array sizes starting from $N=6$, and (iii) are free of any dependencies. 

\textbf{ Motivation}: Given the drawbacks in existing formulations, we propose a three-stage procedure: first, an exhaustive search to find the optimal RMRA configuration for a given sensor count, mainly for $N> 10$; second, using these configurations to determine patterns for algebraic generalizations to fit array descriptions into CFEs; finally, conducting rigorous tests to check if the CFEs are valid for various array sizes.

The following are the key contributions of this work:
\begin{enumerate}
\item We propose a multistage exhaustive search algorithm to solve the combinatorial optimization task of finding optimal RMRAs. A MATLAB code for the same has been provided for reproducibility, making this work the first effort towards formal RMRA synthesis.
\item We found new RMRAs for $N=11$ to $14$, thereby extending the range of known RMRA to sensor counts for $N > 10$.
\item We introduced a new family of suboptimal RMRAs that enjoy CFEs for sensor positions, array aperture, degrees of freedom (DOFs), primary weights, etc. It is also verified that this CFE is valid for any array size of $N\geq8$.
\end{enumerate}

The paper is organised as follows: Section ~\ref{sec:thbck} discusses the fundamentals of the array, Section ~\ref{sec:pm} details the proposed method for building TFRSAs with optimization and closed-form expressions, and Section ~\ref{sec:im} implements the proposed method. Section ~\ref{sec:results} presents the results obtained from the exhaustive search routines. Section ~\ref{sec:cfe} describes the formulation of a new family of TFRSAs obtained through pattern mining on the catalog of arrays discovered through exhaustive searching. Finally, Section ~\ref{sec:end} concludes the paper.

\section{THEORETICAL BACKGROUND}\label{sec:thbck}

This section introduces the fundamental notations and definitions used throughout the paper. Table \ref{glossary} contains the definitions of various parameters of the array, followed by the design rules of TFRSAs. 

\begin{table*}[t]
\centering
\caption{Glossary of Array Processing Terms}
\label{glossary}
\begin{tabular}{>{\bfseries}l p{13cm}}
\toprule
Term & Definition \\
\midrule
Physical Array ($\mathbb{S}$) & The set of sensor positions, normalized to half-wavelengths ($\lambda/2$). \\
Aperture ($L$) & The distance between the first and last sensors in the physical array $\mathbb{S}$. \\
Difference Set ($\mathbb{Z}$) & The set of all pairwise sensor position differences: $\mathbb{Z} = \{ s_i - s_j \,|\, s_i, s_j \in \mathbb{S} \}$. \\
Spatial Lag & An element $z \in \mathbb{Z}$ (any value in the difference set). \\
Difference Coarray (DCA): $\mathbb{D}$ & The set of unique (distinct) spatial lags from $\mathbb{Z}$: $\mathbb{D} = \textit{unique}(\mathbb{Z})$. \\
Holes & Spatial lags that are missing from the difference coarray $\mathbb{D}$. \\
Weight & The frequency (count) of how often a specific spatial lag appears in $\mathbb{Z}$. \\
Degrees of Freedom (DOFs) & The number of distinct entries in the DCA ($|\mathbb{D}|$), quantifying resolvable sources. \\
Weight Function & The set containing the weights for all spatial lags in the DCA. \\
Essential Sensors & A sensor whose removal alters the Difference Coarray (DCA). \\
Fragility & The ratio of essential sensors to the total number of sensors in the array. \\
Signal Model & A model based on the Eigen Value Decomposition (EVD) of the coarray correlation matrix.\cite{re12},\cite{re13},\cite{re14},\cite{re15} \\
\bottomrule
\end{tabular}
\end{table*}

\subsection{TFRSA Design Rules}
A truly robust and valid TFRSA must follow the design principles outlined below.
\subsubsection{\textbf{Healthy Scenario}}
In the healthy configuration, all spatial lags within \([-(L-1), \, (L-1)]\) must have a weight of at least two, and the maximum lag $L$ shall have a weight of one
\begin{equation}
\begin{aligned}
    w(i) &\geq 2, \quad 0 \leq i \leq L-1, \\
    w(L) &= 1
\end{aligned}
\label{healthy}
\end{equation}

The maximum spatial lag $L$ appears exactly once because there is only a single pair of sensors with a separation of $L$ units. It must be noted that the weight function follows even symmetry and therefore the negative weights abide by \(w(-m) = w(m)\).

\subsubsection{\textbf{Single Sensor Failure Scenario}}
The failure of a single sensor (other than $0$ or $L$) must not create gaps in the difference coarray \(\mathbb{D}\). Formally, the weight function of the faulty array shall satisfy the following condition:
\begin{equation}
    w_s(i) \geq 1, \quad 0 \leq i \leq L, \quad \forall s \in \mathbb{S} \setminus \{0,L\}
    \label{failure}
\end{equation}

where $s$ denotes the failed sensor. Equations (\ref{healthy}) and (\ref{failure}) specify the coarray properties and robustness of the array, and are equally instrumental not only for array synthesis but also for array validation.

\section{PROPOSED METHOD}\label{sec:pm}
This section outlines the proposed methodology for obtaining the optimal RMRAs and utilizing the obtained arrays to derive novel closed-form expressions (CFEs).

\tcbset{
  stepbox/.style={
    colframe=purple!80!black,
    colback=white,
    boxrule=0.5pt,
    arc=4pt,
    left=4pt, right=4pt, top=2pt, bottom=2pt,
    width=\linewidth,
  }
}

\begin{figure}[!ht]
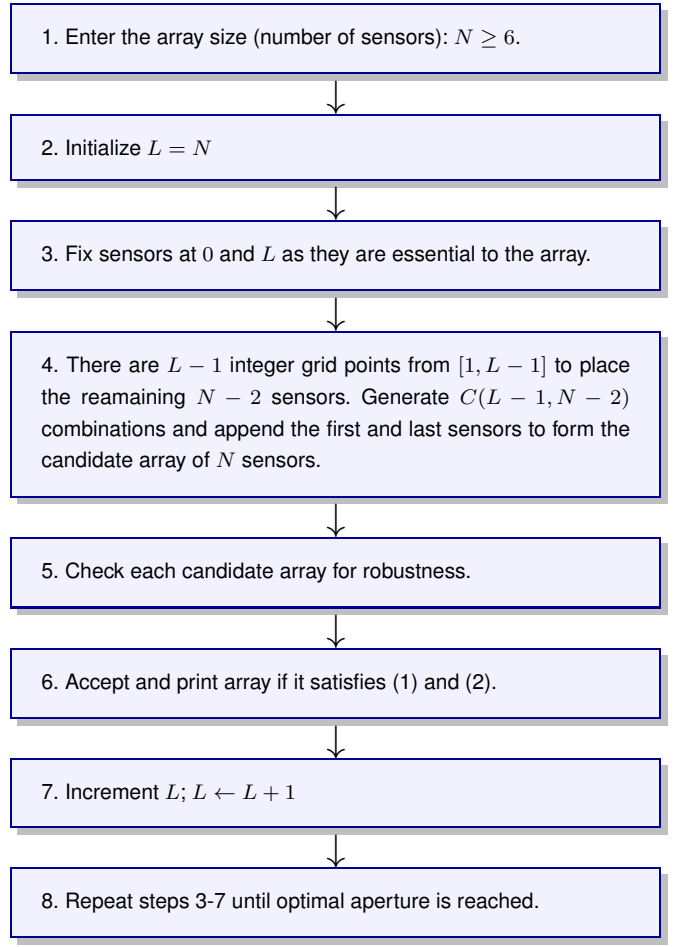

\centering

\tcbset{
  flowstep/.style={
    enhanced,
    colback=blue!5!white,
    colframe=blue!60!black,
    coltitle=black,
    fonttitle=\bfseries,
    sharp corners,
    boxrule=0.8pt,
    top=6pt, bottom=6pt,
    left=8pt, right=8pt,
    fontupper=\sffamily\small,
    colbacktitle=blue!15!white,
    title style={left color=blue!10!white, right color=blue!30!white},
    drop shadow
  }
}

\begin{tcolorbox}[flowstep]
1. Enter the array size (number of sensors): \(N\geq6\).
\end{tcolorbox}

\vspace{-7pt}{\centering\Large$\downarrow$\par}\vspace{-7pt}

\begin{tcolorbox}[flowstep]
2. Initialize $L=N$
\end{tcolorbox}

\vspace{-7pt}{\centering\Large$\downarrow$\par}\vspace{-7pt}

\begin{tcolorbox}[flowstep]
3. Fix sensors at $0$ and $L$ as they are essential to the array.
\end{tcolorbox}

\vspace{-7pt}{\centering\Large$\downarrow$\par}\vspace{-7pt}

\begin{tcolorbox}[flowstep]
4. There are $L-1$ integer grid points from \([1,L-1]\) to place the reamaining $N-2$ sensors. Generate \(C(L-1,N-2)\) combinations and append the first and last sensors to form the candidate array of $N$ sensors.
\end{tcolorbox}

\vspace{-7pt}{\centering\Large$\downarrow$\par}\vspace{-7pt}

\begin{tcolorbox}[flowstep]
5. Check each candidate array for robustness.
\end{tcolorbox}

\vspace{-7pt}{\centering\Large$\downarrow$\par}\vspace{-7pt}

\begin{tcolorbox}[flowstep]
6. Accept and print array if it satisfies (\ref{healthy}) and (\ref{failure}).
\end{tcolorbox}

\vspace{-7pt}{\centering\Large$\downarrow$\par}\vspace{-7pt}

\begin{tcolorbox}[flowstep]
7. Increment $L$; $L\leftarrow L+1$
\end{tcolorbox}

\vspace{-7pt}{\centering\Large$\downarrow$\par}\vspace{-7pt}

\begin{tcolorbox}[flowstep]
8. Repeat steps 3-7 until optimal aperture is reached.
\end{tcolorbox}

\vspace{5pt}
\caption{Flowchart of the Proposed System}
\label{f1}
\end{figure}

\subsection{Optimization Problem}
We formulate the following discrete optimization problem to obtain optimal RMRAs for a known number of sensors.
\begin{equation}
    \begin{aligned}
    &\mathbb{S}_{\text{RMRA}} \triangleq \arg \max_{\mathbb{S}} L \quad \text{subject to} \\ 
        |\mathbb{S}| &= N, \quad |\mathbb{D}_2| = 2L-1\\
        \varepsilon &= {\{0, L\}} 
        \label{optimization}
\end{aligned}
\end{equation}

where, \(|\mathbb{D}_2|=2L-1\) indicates the span of the DCA \([-(L-1), (L-1)]\) where each spatial lag occurs at least twice. The constraint on \(\varepsilon\) forces the requirement of exactly two essential sensors, thereby ensuring functional robustness besides just structural redundancy. 
 
The constraints accurately specify the coarray redundancy and robustness expected of TFRSAs, and the cost function maximizes the array aperture for the given sensor count, indicating that the optimization would result in an optimal RMRA. Although (\ref{optimization}) elegantly represents the NP-hard computational task at hand, its abstraction level is not detailed enough to develop a programming logic, and hence we make use of the constraints in (\ref{healthy}) and (\ref{failure}) as they are more direct and programmer-friendly. This optimization problem will be solved in the MATLAB environment. Fig. \ref{f1} below illustrates the programming logic developed to solve the optimization problem.

\subsection{Procedure to Obtain and Validate the Closed Form Expressions (CFEs)}
Closed-form expressions (CFEs) offer concise mathematical solutions for determining array configurations of any size without exhaustive searches. They enable scalable array designs and are often preferred in sparse array theory, even if they are near- or suboptimal, compared to optimal arrays that require exhaustive searching.

\begin{figure}[!ht]
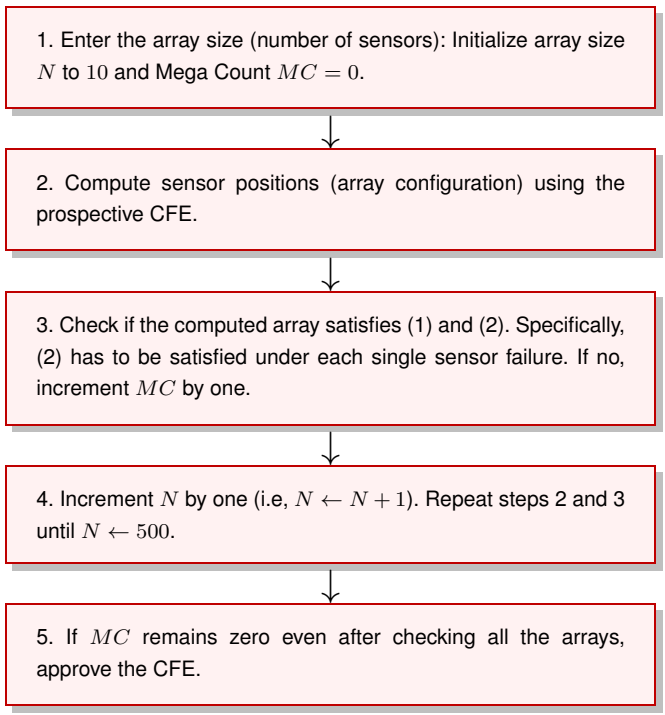

\centering

\tcbset{
  flowstep_red/.style={
    enhanced,
    colback=red!5!white,
    colframe=red!75!black,
    coltitle=black,
    fonttitle=\bfseries,
    sharp corners,
    boxrule=0.8pt,
    top=6pt, bottom=6pt,
    left=8pt, right=8pt,
    fontupper=\sffamily\small,
    colbacktitle=red!15!white,
    title style={left color=red!10!white, right color=red!30!white},
    drop shadow
  }
}

\begin{tcolorbox}[flowstep_red]
1. Enter the array size (number of sensors): Initialize array size $N$ to $10$ and Mega Count $MC = 0$.
\end{tcolorbox}

\vspace{-7pt}{\centering\Large$\downarrow$\par}\vspace{-7pt}

\begin{tcolorbox}[flowstep_red]
2. Compute sensor positions (array configuration) using the prospective CFE.
\end{tcolorbox}

\vspace{-7pt}{\centering\Large$\downarrow$\par}\vspace{-7pt}

\begin{tcolorbox}[flowstep_red]
3. Check if the computed array satisfies (\ref{healthy}) and (\ref{failure}). Specifically, (\ref{failure}) has to be satisfied under each single sensor failure. If no, increment $MC$ by one.
\end{tcolorbox}

\vspace{-7pt}{\centering\Large$\downarrow$\par}\vspace{-7pt}

\begin{tcolorbox}[flowstep_red]
4. Increment $N$ by one (i.e, \(N \leftarrow N+1\)). Repeat steps 2 and 3 until \(N \leftarrow 500\).
\end{tcolorbox}

\vspace{-7pt}{\centering\Large$\downarrow$\par}\vspace{-7pt}

\begin{tcolorbox}[flowstep_red]
5. If $MC$ remains zero even after checking all the arrays, approve the CFE.
\end{tcolorbox}

\vspace{5pt}
\caption{Procedure for obtaining CFEs}
\label{cfe_red}
\end{figure}

The procedure for determining and validating CFEs involves initially creating a catalog of candidate arrays. Pattern mining is then applied to a subset to identify any regular patterns across different sizes or apertures. A prospective CFE is developed based on these patterns. In the second stage, it is verified whether this CFE applies to a wide range of array sizes or only specific N values. The validation process is illustrated in Fig. \ref{cfe_red}.

\tcbset{
  stepbox/.style={
    colframe=purple!80!black,
    colback=white,
    boxrule=0.5pt,
    arc=4pt,
    left=4pt, right=4pt, top=2pt, bottom=2pt,
    width=\linewidth,
  }
}

\section{IMPLEMENTATION METHODOLOGY}\label{sec:im}
This section describes the program developed for the proposed optimization problem, referencing the structured flowchart provided in Fig. \ref{f1}.

\subsection{Initialisation of Parameters}
The user provides the number of array elements, \textbf{$N$}. The program then sets the array aperture, \textbf{$L$}, based on the constraint $L \geq N$, a condition that sets sparse arrays apart from ULAs. A variable, $\mathbf{Z_1}$, is initialized to store the resulting robust arrays. With these parameters set, the program begins generating combinatorial arrays for the specified aperture $L$.

\subsection{Combinatorial Search}
In this phase, for any specified value of \(L\), the program invokes a user-defined function responsible for generating combinatorial arrays. This function accepts three parameters: the initial element of the array, designated as \(0\); the terminal element, specified as \(L\); and the total number of elements in the array, denoted as \(N\). Consequently, it produces an output containing the combinatorial arrays. Once the inputs are established, an array spanning from \(0\) to \(L\) is constructed, and all possible combinations of size \(N\) are generated and stored. Subsequently, any duplicate entries within the cell array are eliminated, and the resulting set of combinatorial arrays is returned to the main program.

\subsection{Array Robustness Validation}
To ensure robustness against single-sensor failure, arrays undergo a two-step validation. In the first stage, it is checked if the array satisfies (\ref{healthy}). If yes, the array is retained for further testing. In the second stage, arrays are tested for robustness by simulating the removal of each sensor individually. In each case of single sensor failure, the array shall satisfy (\ref{failure}).

All validated arrays are stored in a set $\mathbf{Z_1}$, which is then sorted and cleared of duplicates. The search process iterates through increasing apertures, terminating when no valid array is found for a particular $L+x$. The program then presents all valid TFRSAs generated for apertures up to $L+x-1$. The code for this method is available at: \textbf{\href{codelink}{https://doi.org/10.5281/zenodo.17087885}}.

\subsection{MEX Code Generation}
MEX files are pre-compiled C/C++ programs derived from MATLAB code, offering faster execution akin to machine-level performance compared to .m files, especially for larger values of $N$. To enhance efficiency in tasks like Combinatorial Search and Array Robustness Validation, MEX versions are developed, significantly reducing the time for array generation and testing. MATLAB's Coder tool facilitates the creation of MEX files by allowing function selection, input data type specification, and testing, ultimately producing a faster executable directly usable in programs.

\section{RESULTS}\label{sec:results}
The MATLAB program successfully generated optimal RMRAs for known sensor counts, confirming its functional correctness. It was then used to explore RMRAs beyond the known range.

\subsection{Optimal Arrays for \({N\leq 10}\) Sensors}\label{AA}
In this section, we discuss the results achieved through the generation of optimal arrays for \(6 \leq N \leq 10\), targeting the maximum possible aperture. The obtained results were in agreement to the optimal RMRA configurations published by Liu and Vaidyanathan in the past \cite{re5}.

\subsection{Optimal Arrays for Sensor Counts ${N = 11}$ to ${14}$}
The current discussion pertains to the generation of arrays for the values of \(N = 11\) to \(14\). Consider the case where \(N=11\). The process commences with \(L=11\), indicating that the algorithm must select intermediate values from the set \([1,10]\), with the endpoints 0 and 11 being fixed. The total number of combinations in this scenario is \(C(10,9) = 10\). Upon identifying a valid solution, the algorithm increments \(L\) by 1 and reiterates the process. At each subsequent stage, the number of potential combinations increases, as demonstrated in Table \ref{tab1114a}.

\begin{table}[htbp]
\caption{Comparison of Combinatorial Search Space Before and After Fixing Sensors at Mandatory Positions}
\label{tab1114a}
\centering
\begin{tabular}{|c|c|c|l|l|}
\hline
\textbf{Stage} & \textbf{$L$} & \textbf{Range} & \textbf{Before Fixation} & \textbf{After Fixation} \\
\hline
1  & 11 & [0, 11] & $C(10,9) = 10$      & $C(7,6) = 7$ \\
2  & 12 & [0, 12] & $C(11,9) = 55$      & $C(8,6) = 28$ \\
3  & 13 & [0, 13] & $C(12,9) = 220$     & $C(9,6) = 84$ \\
4  & 14 & [0, 14] & $C(13,9) = 715$     & $C(10,6) = 210$ \\
5  & 15 & [0, 15] & $C(14,9) = 2002$    & $C(11,6) = 462$ \\
6  & 16 & [0, 16] & $C(15,9) = 5005$    & $C(12,6) = 924$ \\
7  & 17 & [0, 17] & $C(16,9) = 11440$   & $C(13,6) = 1716$ \\
8  & 18 & [0, 18] & $C(17,9) = 24310$   & $C(14,6) = 3003$ \\
9  & 19 & [0, 19] & $C(18,9) = 48620$   & $C(15,6) = 5005$ \\
10 & 20 & [0, 20] & $C(19,9) = 92378$   & $C(16,6) = 8008$ \\
11 & 21 & [0, 21] & $C(20,9) = 167960$  & $C(17,6) = 12376$ \\
12 & 22 & [0, 22] & $C(21,9) = 293930$  & $C(18,6) = 18564$ \\
13 & 23 & [0, 23] & $C(22,9) = 497420$  & $C(19,6) = 27132$ \\
\hline
\end{tabular}
\end{table}
At stage $13$, the algorithm exhaustively evaluates all the 497,420 possibilities but fails to find a valid solution and, therefore, returns the last successful solution found at stage $12$, with $L=22$ as the optimal array.

We also observed that the first three and the last two sensors in the array are important to preserve the multiplicity of the corner lags $L$, $L-1$, and $L-2$. Consequently, the initial three and final two elements of the array can be fixed, i.e., $0,1,2,L-1,L$, so that the modified inputs for the combinatorial search would be $3, L-2$ and $N-5$. By fixing these elements, the total number of array combinations to be generated is reduced from \(C(L-1, N-2)\) to \(C(L-4, N-5)\). This reduction in possible combinations considerably narrows the search space, thereby facilitating faster computational performance, as shown in the last column of Table 1.

Table \ref{tab1114} shows arrays with maximum apertures for \(11\leq N \leq 14\), specifically $22, 26, 32$, and $36$, which are sparse. Fixing elements minimized computational time and overload without compromising program validity, thus improving efficiency. 

\begin{table}[htbp]
\caption{Optimal Arrays for $N = 11$ to $14$}
\label{tab1114}
\centering
\begin{tabular}{|c|l|}
\hline
\textbf{N} & \textbf{Array Configuration} \\
\hline
11 & [0\quad 1\quad 2\quad 3\quad 4\quad 10\quad 11\quad 16\quad 17\quad 21\quad 22] \\
12 & [0\quad 1\quad 2\quad 3\quad 4\quad 5\quad 12\quad 13\quad 19\quad 20\quad 25\quad 26] \\
13 & [0\quad 1\quad 2\quad 4\quad 5\quad 9\quad 14\quad 19\quad 24\quad 25\quad 30\quad 31\quad 32] \\
14 & [0\quad 1\quad 2\quad 3\quad 4\quad 5\quad 12\quad 14\quad 21\quad 23\quad 29\quad 30\quad 35\quad 36] \\
\hline
\end{tabular}
\end{table}

After $N=14$, when the program attempted to compute optimal arrays for $N=15$, it encountered an error during execution due to the \textit{combinatorial explosion} of candidate solutions. As the array aperture $L$ increases, the number of sensor placement combinations, given by \(^{L-4}C_{\, N-5}\), explodes. For moderate values of $N$ and $L$, this leads to memory and computational limits in MATLAB thereby causing the following error:
\texttt{Cannot grow a matrix with greater than intmax() elements}, is 
triggered by the \texttt{nchoosek} function. Even with MEX-based acceleration, enumerating all possible sensor placements becomes infeasible due to exceeding integer and memory bounds. This limitation restricts the scalability of exhaustive search methods and motivates the need for alternative optimization-based or heuristic approaches.

\subsection{Near Optimal Arrays for \(N\geq15\) }
After generating arrays up to $N = 14$, we began examining arrays with $ N \geq 15$ under certain constraints. In this section, we discuss the findings related to near-optimal arrays with the largest possible apertures. Some of these findings are summarized in Table \ref{tab1520}, provided below.

\begin{table}[htbp]
\caption{Near-Optimal Arrays for $N = 15$ to $20$ with a Few Fixed Sensor Positions on Either End}
\label{tab1520}
\centering
\scriptsize 
\begin{tabular}{|c|p{5.8cm}|c|c|}
\hline
\textbf{N} & \textbf{Array Configuration} & \textbf{First} & \textbf{Last} \\
\hline
15 & [0 1 2 3 5 8 12 19 25 26 31 35 39 40 41] & 4 & 3 \\
16 & [0 1 2 3 13 15 17 19 35 36 39 40 43 44 45 46] & 4 & 4 \\
17 & [0	1 2	3 4	5 22 27 32 33 38 39 44 45 46 47 48] & 5 & 5 \\
18 & [0	1 2 3 4 7 9 16 24 32 34 40 42 50 51 52 53 54] & 5 & 5 \\
19 & [0	1 2 3 4 5 6 26 32 38 39 45 46 52 53 54 55 56 57] & 6 & 6 \\
20 & [0	1 2	3 4	5 7 12 19 20 27 35 48 56 57	58 59 60 61	62] & 6 & 7 \\

\hline
\end{tabular}
\end{table}

To manage the intensive combinatorial search for large arrays, a brute-force method with optimizations is used: fixing elements to reduce the search space and early termination once a set number of valid arrays is found. This reduces combinations significantly, as shown for $N=16$ from $C(47,16)$ to $C(39,8)$ and for $N=20$ from $C(63,20)$ to $C(50,7)$. The method balances computational efficiency with solution optimality, where more fixed elements speed up processing but may result in suboptimal arrays.

This method, in conjunction with the MEX code technique discussed in Section IV, is an innovative aspect of this program that decreases computational load.
\subsection{Application of 2-FRSAs in Radar Systems}
2-fold redundant sparse arrays, including nested and co-prime configurations, are crucial in radar technology beyond direction finding, aiding in imaging, DoA estimation, and SAR. They create a hole-free difference coarray with $O(N^2)$ virtual DoF using $N$ elements, enabling high angular resolution for resolving close targets in automotive radar and electronic warfare. This redundancy supports spatial smoothing for robust subspace algorithms like MUSIC. In radar imaging, it enhances cross-range resolution for detailed images with fewer channels, reducing cost and complexity. MIMO-SAR systems use these arrays for ultra-high-resolution mapping, minimizing physical footprint and improving parameter estimation by reducing mutual coupling effects.

\section{A NOVEL SUB-OPTIMAL TFRSA FAMILY WITH CLOSED-FORM FORMULATIONS}
\label{sec:cfe}
By closely observing the sensor positions for $N=11$ and $N=12$ in the Table. \ref{tab1114}, we formulated the following closed-form expressions (CFEs). The proposed sub-optimal RMRA configuration can be characterized by the algebraic CFE given below:
\begin{equation}
\mathbb{S} = 
\left\{
\begin{array}{l}
\underbrace{0, 1, 2, \ldots, p {-} 1}_{\text{ULA segment of $p$ sensors}}\\[12pt]
\underbrace{(2p), (2p {+} 1),(3p {+} 1), (3p {+}2), (4p {+} 1), (4p {+} 2)}_{\text{Last six sensors that lead to sparsity}}
\end{array}
\right\}
\label{pmcfe}
\end{equation}

where, \(p=N-6\). This is valid for $N\geq8$ (as proved shortly). The array can also be expressed using the IES notation given below:
\begin{equation}
\mathbb{I}_{\text{RMRA}} = \{ 1^{p-1},\ (p+1),\ 1,\ p,\ 1,\ (p-1),\ 1 \}
\label{eq2}
\end{equation}
This CFE was tentatively validated using the procedure outlined in Fig. \ref{cfe_red}.
\subsection{\textbf{Array Aperture}}
As known, the separation between the corner sensors of the array defines the array aperture. As the last sensor lies at $4p+2$ and the first sensor lies at the origin, the array \(\mathbb{S}\) has an aperture $L$ given by:
\begin{equation}
    L = 4p+2 = 4N-22
\label{eq3}
\end{equation}
Based on ~\eqref{eq3}, one can determine the smallest $N$ that satisfies  ~\eqref{pmcfe}. The array shall provide at least one unit of aperture higher than that of the ULA to qualify as a sparse array. Hence, we have the inequality $4N-22 > N-1$, which results in $N>7$. Hence, the proposed array can be constructed if there are eight or more sensors \((N\geq8)\).

\subsection{\textbf{DCA Span and DOFs Offered}}
As the proposed array is hole-free from \([-L,L]\), the DCA span $\mathbb{D}$ also denotes the uniform DOFs offered by the array and is given by
\begin{equation}
    \mathbb{D} = 2L+1 = 8N-43  
    \label{eq4}
\end{equation}
As such, the array provides $O(N^2)$ DOFs for $N$ sensors.

\subsection{\textbf{Representative Examples of the Proposed Array}}

\subsubsection{\textbf{Smallest Array Satisfying the Proposed CFE}}
\begin{figure}[ht]
        \centering
        \includegraphics[width=\linewidth]{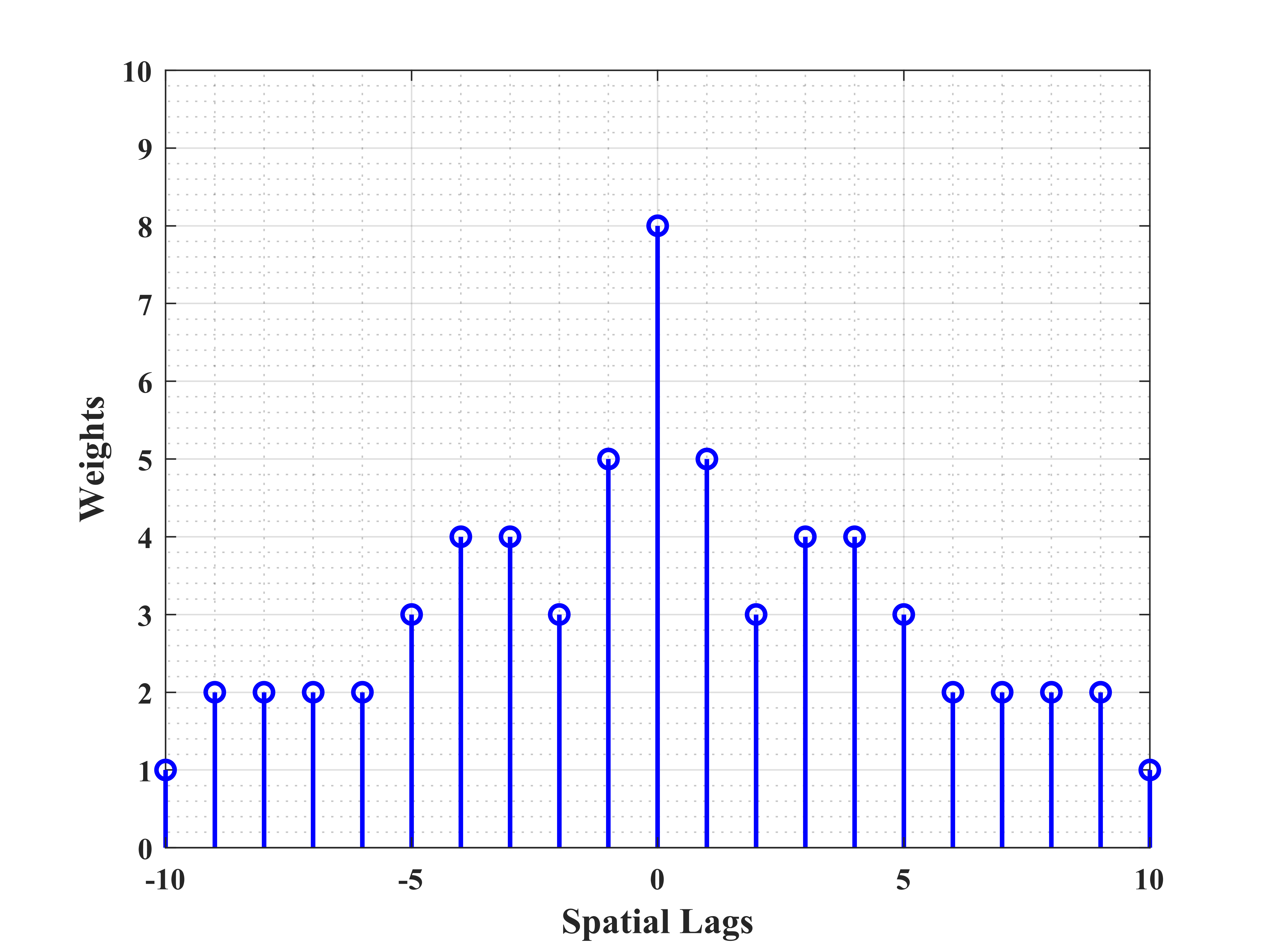}
        \caption{Weight Function in Healthy Case}
        \label{fig:healthyf}
    \end{figure}

\begin{figure}[ht]
        \centering
        \includegraphics[width=\linewidth]{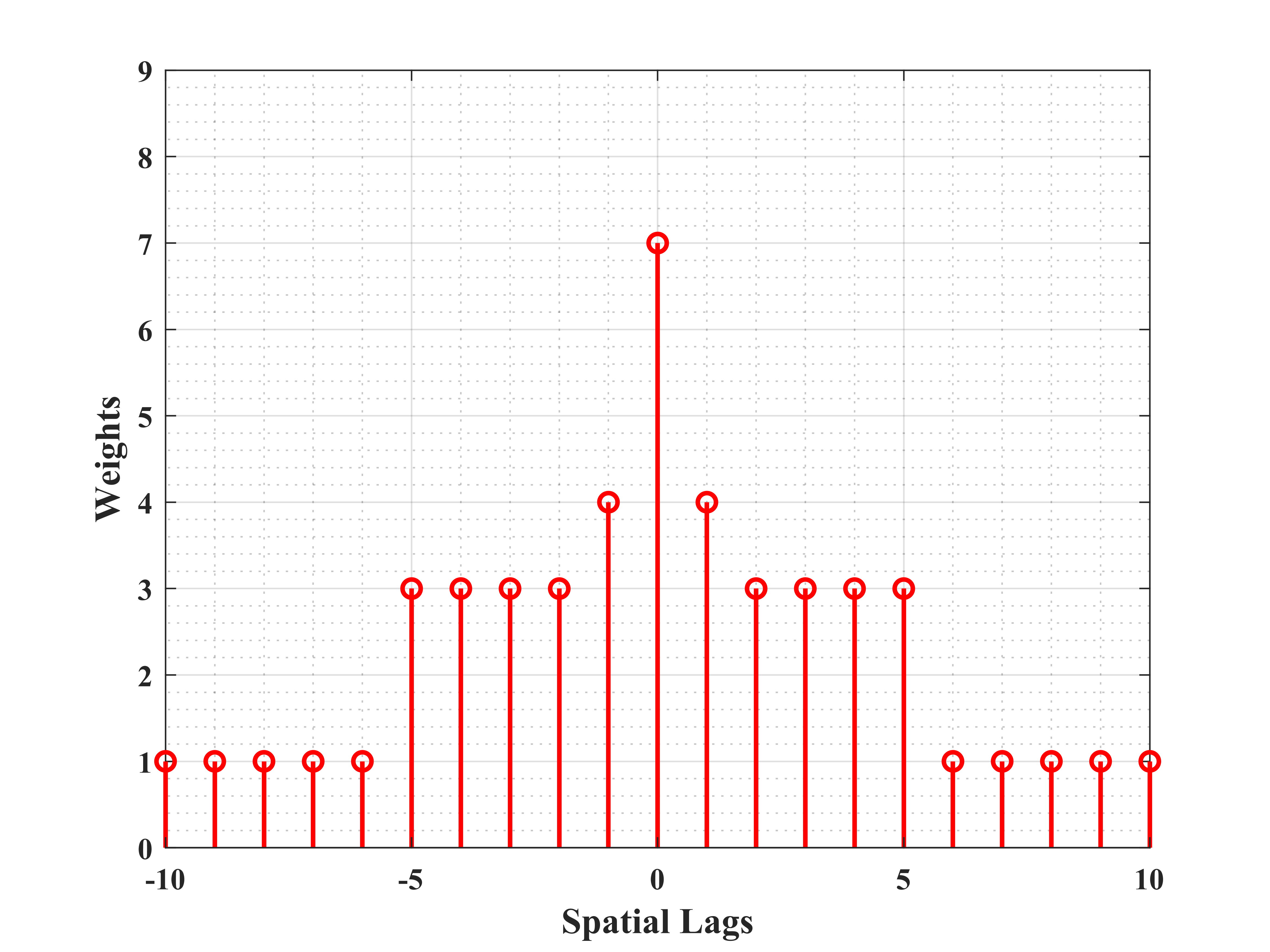}
        \caption{Weight Function in Faulty Case: Sensor Failure at Position 1}
        \label{fig:failure1f}
\end{figure}

As mentioned above, the proposed array is valid for all array sizes from $N=8$. Substituting $N=8$ i.e. $p=2$ in (\ref{pmcfe}), we get the array configuration \([0, 1, 4, 5, 7, 8, 9, 10]\). Figs.~\ref{fig:healthyf} and ~\ref{fig:failure1f} shows the weight function of this array during healthy and faulty cases. As seen from Fig.~\ref{fig:healthyf}, all the weights from \([-9, 9]\) are at least two, indicating that the array has at least two sensor pairs to generate each of these lags. Fig.~\ref{fig:failure1f} shows the weight function of the faulty array when the sensor at position \({1}\) fails. It can be seen that all weights from \([-10, 10]\) are at least one, indicating a hole-free DCA. A similar analysis can be repeated by failing each sensor at a time and verifying the respective weight function. Because thorough failure analysis using the above procedure can get cumbersome for large arrays, we used the following program to automate the failure analysis and to get a comprehensive summary of the array's robustness (\textbf{{https://github.com/profashish/two-fold-robustness-check}}). 


\subsubsection{\textbf{Array Configuration for any arbitrary $N$}} 
Using the proposed CFE in (\ref{pmcfe}), one can instantly obtain the array configuration (sensor positions) for any $N$ without the need for exhaustive searching. For example, the array configuration for $N=100$ will be \([0, 1, 2,..., 93, 188, 189, 283, 284, 377, 378]\). Although this CFE leads to a sub-optimal aperture, it provides a tangible means to realize large arrays, thereby making the design instantly scalable and computationally tractable. 

\subsection{\textbf{Comparison with Existing TFRSAs}}
\begin{table*}[t]
\centering
\caption{Comparison of Existing TFRSA Formulations}
\label{tab:array_comparison}
\renewcommand{\arraystretch}{1} 
\setlength{\tabcolsep}{6pt}       
\small
\begin{tabular}{|l|c|c|c|c|c|}
\hline
\textbf{Criteria} & \textbf{Symmetric Nested Array} & \textbf{Composite Singer Array} & \textbf{RMRA} & \textbf{2FRA (Zhu et al.)} & \textbf{Proposed} \\
\hline
Availability of CFEs & \checkmark & \checkmark & $\times$ & \checkmark & \textbf{\checkmark} \\
\hline
CFEs valid for all array sizes $N \geq 10$ & $\times$ & $\times$ & N/A & \checkmark & \checkmark \\
\hline
Free from hidden dependencies & \checkmark & \checkmark & \checkmark & $\times$ & \textbf{\checkmark} \\
\hline
\end{tabular}
\end{table*}

It can be seen from Table \ref{tab:array_comparison} that the proposed array overcomes all the drawbacks in existing TFRSAs. Although the 2FRAs proposed by Zhu et. al. are elegant \cite{re6}, they suffer from hidden dependencies as explained in \cite{re11}. The only drawback of the proposed arrays is that they provide only \(O(N)\) DOFs for $N$ sensors as opposed to the \(O(N^2)\) DOFs provided by 2FRAs. Moreover, the proposed array is highly susceptible to mutual coupling due to its structural similarity to the ULA. Nevertheless, lower DOFs and higher mutual coupling are not so severe problems as having hidden dependencies which can create unexpected uncertainties in the DOA estimation performance of the array.

\section{CONCLUSION AND FUTURE SCOPE}
\label{sec:end}
A MATLAB program was developed to generate optimal RMRAs by taking the number of sensors $N$, as the input parameter. The algorithm constructs combinatorial arrays and checks if they satisfy the two constraints of offering a double difference base during the healthy case and hole-free DCA during the failure of a single sensor. The search process proceeds across multiple stages and terminates when the optimal aperture is reached. Measures such as MEX code implementation and prefixing sensors at designated positions have been employed to speed up the program execution. Optimal RMRAs for $N=11$ to $N=14$ were found using the developed program, thereby extending the catalog of known RMRAs. A major finding from this work is the inability of MATLAB's \texttt{nchoosek} function to store large rows of number combinations due to the intmax limitation. Additionally, hidden and repetitive patterns in the sensor placement across various array sizes was analyzed and closed-form expressions were constructed to specify a new family of sub-optimal RMRAs. 
\subsection{Limitations}
Although the proposed multistage exhaustive search has no flaws, the MATLAB implementation has limitations that hinder its scalability to large arrays. As noted in Section V-B, a data structure issue prevented the program from functioning normally for $N \geq 15$. We are currently exploring alternative approaches to overcome this bottleneck. Additionally, the primary weights of the proposed suboptimal RMRA is highly susceptible to mutual coupling. This is due to the large ULA segment consisting of $N-6$ sensors at the beginning of the array. 
\subsection{Future Scope}
In the future, systematic methods such as Branch and Bound (B\&B) or mixed integer linear programming (MILP) could be explored to solve the NP-hard computational task. At the same time, alternate platforms such as Gurobi/CPLEX and programming languages such as Python will be explored to overcome the data structure limitation encountered in MATLAB. The overall aim would be to determine optimal RMRAs till $N=20$ and near-optimal RMRAs from $N=21$ to $N=30$ to extend the catalog of known RMRAs. Subsequently, the extended catalog could be utilized to derive newer CFEs for TFRSAs/near-optimal RMRAs with higher immunity to mutual coupling and/or higher DOFs \(O(N^2)\) than existing solutions. Such robust arrays could find applications in sparse Multiple-Input Multiple-Output (MIMO) automotive radar arrays to ensure accurate and fault-tolerant angle estimation \cite{re17},\cite{re18}, \cite{re19}. MIMO-Synthetic Aperture Radar (SAR) systems could make use of these arrays for ultra-high-resolution mapping and robust parameter estimation even during single sensor failures.

\end{document}